# Convergence and cluster structures in EU area according to fluctuations in macroeconomic indices

**Mircea Gligor**[1, *]

[1] National College "Roman Voda" Roman-5550, Neamt, Romania, Euroland
e-mail: mrgligor@yahoo.com
* Corresponding author

**Marcel Ausloos** [2]

[2] GRAPES, B5, Sart Tilman, University of Liege, Belgium, Euroland
e-mail: Marcel.Ausloos@ulg.ac.be

**Abstract:** The cluster analysis methods are used in order to perform a comparative study of 15 EU countries in relation with the fluctuations of some basic macroeconomic indicators. The statistical distances between countries are calculated for various moving time windows, and the time variation of the mean statistical distance is investigated. The decreasing of the mean statistical distance between EU countries is reflected in the correlated fluctuations of the basic ME indicators: GDP, GDP/capita, Consumption and Investments. This empirical evidence can be seen as an economic aspect of globalization. The Moving Average Minimal Length Path (MAMLP) algorithm allows to search for a cluster-like structures derived both from the hierarchical organization of countries and from their relative movement inside the hierarchy. It is found that the strongly correlated countries with respect to GDP fluctuations can be partitioned into stable clusters. Some of the highly correlated countries, with respect to GDP fluctuations, display strong correlations also in the Final Consumption Expenditure, while others are strongly correlated in Gross Capital Formation. On the other hand, one notices the similitude of the classifications regarding GDP and Net Exports fluctuations as concerns the squared sum of the correlation coefficients (so called country "sensitivity"). The final structure proves to be robust against the constant size time window moving over the scanned time interval. The policy implications of the above empirical results concern the economic clusters arising in the presence of Marshallian externalities and the relationships between trade barriers, R&D incentives and growth that must be accounted in elaborating a cluster-promotion policy.

# 1   Introduction

The problem of studying the economic growth patterns across countries is actually a subject of great attention to economists. An important reason for the increasing interest in this problem is that "persistent disparities in aggregate growth rates across countries have, over time, led to large differences in welfare" (Durlauf and Quah, 1999). The intellectual payoffs of comparative studies may be high: moreover various patterns of growth can be inferred from the statistical data, the statistical methodology itself might be considerably enriched.

On the other hand, it is well known that a general question facing researchers in many areas of inquiry is how to organize observed data into meaningful structures, that is, to develop taxonomies. In this sense, cluster analysis is an exploratory data analysis tool which aims at sorting different objects into groups in a way that the degree of association between two objects is maximal if they belong to the same group and minimal otherwise. The term "cluster analysis" (first used by Tryon, 1939) refers to a number of different algorithms and methods for grouping objects of similar kinds into respective categories. The paper is built upon these two considerations.

Consider first the two groups of issues of actually increasing interest in economic growth literature: the first refers to the economic *convergence* of countries and regions, while the second pertains to the country *differentiation*, or *clustering*, as a result of the disparities in their growth rates.

(I) As regards to the first sort of issues, it is of interest to examine whether the economic convergence of EU-15 countries may be empirically argued starting from the time evolution of the basic macroeconomic indicators. Moreover, whether this phenomenon (in so far as it does) occurs continuously or intermittently, and what is the role the time window size in studying it; another point is whether the phenomenon may be related to the emergence of cooperation in social/ecological systems;

(II) Concerning the second sort of issues, it is worth to call in question the most appropriate methodology from which a robust country clustering structure can be derived and if this cluster-like structure has any economic support; moreover, it would be of interest to investigate the possible connections between the country clustering and the speciation in ecological/ biological systems.

The economic convergence has a particular place in the increasing literature of economic growth during the last few years. The OECD Economic Survey of the Euro Area (2004) promoted the idea of the convergence in economic development as a prime policy goal of the European Union. The same document includes observations such as "Per capita GDP has tended to converge between countries, but evidence of convergence across regions is mixed" and "this slow pace of convergence may partly reflect the timid pace of integration, while the evolution of human and physical capital endowments was uneven across countries and regions". These findings seem to plead for a European cluster-like structure rather than for a European convergence.

Practically the problem of "countries convergence" is usually addressed from two different viewpoints: (1) business cycle synchronisation and (2) so called $\sigma$-convergence.

(1) There is now a large literature that examines different questions related to the extent of synchronisation of the international business cycle. The correlations in the post-war period seem to support the idea of regional cycles, rather than the one of a common international cycle. For example, Backus and Kehoe (1992) found that German cycles are significantly positively correlated with Italian and UK cycles for example, while Canadian and US cycles are also highly positively correlated. As regards the European area, Artis and Zhang (1999) argued that European integration and associated Exchange Rate Mechanism have produced a region-specific European business cycle that has become more synchronized around the German business cycle and less attached to the US cycle, while Frankel & Rose (1998) suggested a strong relationship between trade linkages and cycle synchronicity. In the same idea Inklaar and de Haan (2001) showed that the relationship between exchange rate stability and business cycle synchronisation can be broken once different sub-periods are analysed. Recently, Bodman and Crosby (2005) have found that "in general one could reject the null of independent recession dates in the G7 countries. Overall, these rejections are consistent with an interpretation of regional synchronisation".

(2) On the other hand, the economic growth literature often resorts to the concepts of σ and β-convergence, first introduced in Sala-i-Martin (1990). The β-convergence, a concept emerging from neo-classical growth models assuming diminishing returns in production, refers to a potentially negative relationship between growth in per capita GDP and the initial level of income of a country, so that poorer countries may grow faster than richer countries, and thereby catch up with these richer countries. In contrast the concept of σ-convergence is related to the income distribution of a set of economies. In fact, the existence of σ-convergence implies that the world income distribution shrinks over time. Thus, for example, if we consider the variance (or the standard deviation) of the log of GDP at a certain time $t$ and at time $t + T$ ($T > 0$), we say that there is σ-convergence for a given set of economies and for a given period of time ($T$), if: $\sigma^2(t) > \sigma^2(t + T)$. A number of studies have aimed to test empirically whether β-convergence has been observed. While initial studies reported a certain (small) rate of convergence (e.g., Barro, 1991; Sala-I-Martin, 1996), more recent research has put these initial findings in doubt (Caselli et al., 1996; Bliss, 1999; Cannon and Duck, 2000). More recently, Furceri (2005) as well as Wodon and Yitzhaki (2006) demonstrated that σ-convergence is only a sufficient (but not necessary) condition for the existence of β-convergence.

In spite of the increasing number of papers pertaining to country comparative studies literature, there are relatively few authors inclined to embody in their methodological arsenal the recent developments in the "exotic" fields such as graph theory, hierarchical networks and cluster analysis. We have to mention here several remarkable exceptions (Quah, 1996; Hill, 2001; Andersen, 2002; Mora et al, 2005 among others), part of them playing the role of underlying incentives for us in elaborating the present study.

To avoid turning our paper into a technical-oriented one, or worse, falling in a futile exercise in data mining, we address at this point the question whether *the cluster-like structure has some support in the present economic literature*.

Growth literature often considers the existence of groups of economies which have been termed "convergence clubs" that present a homogeneous pattern and converge towards a common steady state. In the endogenous theoretical framework suggested in Azariadis and Drazen (1990) externalities could explain the presence of spatial regional clusters that share lower or higher levels of development. Empirically, Chatterji (1992) detected two convergence clubs for a sample of 109 countries, the US being the leader. At the same time, Ben-David (1994) proposed local convergence, dividing world economies into three groups, among which the poorest is also the largest. Quah (1996), (1997), proposed two approaches in order to explain the existence of convergence clubs: an endogenous formation of coalitions, and the generation of several dynamics of convergence that depend on the initial characteristics of the distribution. In his approach, richer regions tend to converge towards a middle rich position, whereas poorer ones tend to a middle poor position. Convergence may then be maintained inside clusters but not between them (Durlauf and Quah, 1999). Mora (2005) considered the possibility that European regional economies could be classified into different convergence clubs, considering optimum criteria of minimizing the loss of information when groups are configured.

There is also a large support for apparently industrial clustering. According to Krugman (1991); Fujita et al. (1999) among others, the concentration of industrial activities across space is primarily influenced by historical accidents. Instead, Barrios and Strobl (2004) studied the pattern of geographic concentration of industries in EU countries and regions between 1972 and 1995 and conclude that "the observed rise in concentration of manufacturing activities is generally due to randomness in the distribution of countries' and regions' industrial growth, a feature which has not been yet considered by the empirical literature concerning the European case". The problem of industrial clustering is often associated to the one of the common patterns in the firm growth dynamics (Giuliani et al., 2005; Mehrotra and Biggeri, 2005; Yeung et al., 2006).

A cluster-like structure may be also derived from the consumption patterns. When trade patterns between nations are modelled as general equilibrium allocations between risk-averse trading partners, a high correlation of consumption across countries is involved. Although the data analysed by Backus et al. (1992) showed a clear tendency for cross-country output correlations to be higher than cross-country consumption correlations, Pakko (2004) performed a spectral decomposition of the consumption /output correlation puzzle and showed that the above finding holds "only within the range of frequencies generally associated with business cycle fluctuations. At both higher and lower frequencies, cross-country consumption correlations show a greater tendency to exceed output correlations".

To consider that convergence is proved through the decrease of the mean statistical distance among countries by means of their annual rates of growth, without taking into account their initial level of development, implies that *only* σ-convergence may be relevant. Moreover, while it has been recently shown that β-convergence can be observed both forward and backward in time (Wodon and Yitzhaki, 2006), in this approach the concept of convergence appears closely related to the time arrow and to the presence of exogeneus or endogenous shocks. So, it aquires the features of an *adaptive* processus, in the same sense as the adaptive emergence of cooperation occurs in ecological systems.

Indeed, the evolution of cooperation and collective action catches more and more attention in the economics framework. Most models and experiments have been pursued in a game-theoretic context and involve some payoffs as reward or punishment (Lewontin, 1961; Maynard Smith, 1982, and others). More recently, Durrett and Levin (2005) have shown that these payoffs are unnecessary, and that stable social groups can sometimes be maintained provided simply that the agents are prone to imitate each others. On the same way, Horan et al (2005) have gone further, showing how the endogenous division of labour and subsequent trading among early modern humans could have helped them to survive.

However, as we indicated in the 2nd paragraph of this Introduction, the second sort of issues calls into question the appropriateness and limitations involved by using the minimal spanning tree (MST) and other similar cluster-deriving algorithms in the macro-economic framework.

As one might search for a cluster-like structure based on the strongest correlations and anti-correlations between time series, it is appropriate to recall other classification tree methods in statistics. Long ago, methods as CHAID (Chi-squared Automatic Interaction Detector) proposed by Kiss (1980), the classical C&R Trees (Classification and Regression Trees) Algorithm (Breiman et al., 1984) and other tree classification techniques have been discussed. They are known to have a number of advantages over many other techniques. In most cases, the interpretation of results summarized as on a tree is very simple. This simplicity is useful not only for purposes of rapid classification of new observations, but can also often yield a simple "model" for explaining why observations are ordered or predicted in a particular manner. On the other hand, the final results of using tree methods for classification or regression can be summarized in a series of (usually few) logical if-then conditions (tree nodes). Therefore, there is no need of an implicit assumption on the underlying relationships between the predictor variables. Thus, tree methods are particularly well suited for data mining tasks, when there is no coherent comprehensive theories regarding which variables are interrelated or how.

The above considerations (among many other similar ones) suggest a large support for various kinds of *taxonomies* at different levels of the economic activity. One can recall here that taxonomies are of common use in biology, physics, and computer sciences as well as in other various fields; it is useful to adopt from these so "convergence" in methodology. The next section of the paper may be seen as intended for that purpose.

A tree clustering method uses the dissimilarities (similarities) measured as distances between objects when forming the clusters. Therefore, in tree-like classifications, the first problem is to choose an adequate distance measure in order to place progressively greater weight on objects (say series $\{x_i\}$ and $\{y_i\}$) that are further apart.

Various definitions of distances are proposed in the statistics literature so far. We recall here only those of common use, as the Euclidean distance:

$$d(x, y) = \left( \sum_i (x_i - y_i)^2 \right)^{1/2} \tag{1}$$

and the City-block (Manhattan) distance:

$$d(x, y) = \sum_i |x_i - y_i| \tag{2}$$

The first definition has a few advantages, e.g., the distance between any two objects is not much affected by the addition of new objects in the analysis, which may be outliers. The distance (1) can be generalized as a "power distance":

$$d(x, y) = \left( \sum_i (x_i - y_i)^p \right)^{1/r} \tag{3}$$

where $p$ and $r$ are user-defined parameters, or as a correlation (statistical) distance:

$$d(x, y) = [2(1 - C(x, y))]^{1/2} \qquad (4)$$

where the $C$ is the correlation coefficient:

$$C(x, y) = \frac{<x_i y_i> - <x_i><y_i>}{\sqrt{<x_i^2> - <x_i>^2><y_i^2> - <y_i>^2>}} \qquad (5)$$

As a matter of fact, we have into view a classification-type problem that is to predict values of a categorical dependent variable (class, group membership, etc.) from a predictor variable which is - in our approach - the correlation coefficient.

As we aim to search for a country hierarchical structure starting from the correlations between several time series describing their macroeconomic evolution, the statistical distance (1.4) is used in the present approach, though we admit that other choices could be of interest[1].

The method used here below, namely the moving-average-minimal-length-path (MAMLP) is described in Section 2, with other several related techniques. In essence, MAMLP was derived by applying the minimal-length-path-to-average classification to various moving time windows. In other words, as a first step, for each time window a hierarchy of countries was found taking their minimal path distance on average; thereafter, in a second step the strongest correlations and anti-correlations between the movements of countries inside the hierarchy were investigated.

The considered macroeconomic indicators are GDP, GDP/capita (GDPC), Final Consumption Expenditure (FCE), Gross Capital Formation (GCF) and Net Exports (NEX).

The results are presented in Section 3. Firstly, the data sources are presented. Then, this section groups the results in relation with the multiple aims of our investigation: first, the relevant role of the time window size is pointed out by studying GDP/capita in two moving time windows of 10 and 5 years sizes respectively; secondly, GDP is investigated in a moving time window of 5 years, and the MAMLP method is applied to find the strongest correlations and anti-correlations between countries, which result in a cluster-like structure; thirdly, the same method is applied to the other three indicators (FCE, GCF and NEX), which are usually considered as basic ingredients in the GDP estimation.

Conclusions are found in Section 4. A statistical test of robustness, namely the shuffled data analysis, is done in Appendix 1; the tables of MAMLP distances and corresponding correlation matrices for FCE, GCF and NEX are given in Appendix 2, while a possible extension to a multivariate approach, namely the Cluster Variation Method, is done in Appendix 3.

---

[1] For example, there has been some recent interest in extending the idea of distance or dissimilarity between two objects to that of triadic distances between three objects (Daws, 1996; Heiser and Bennani, 1997). The triadic distances are usually defined as functions of the pair-wise or dyadic distances (de Rooij and Heiser, 2000). More recently, Gower and de Rooij (2003) demonstrated that the multidimensional scaling of triadic distances (MDS3) and the conventional one of dyadic distances (MDS2) both give Euclidian representations and can be expected to give very similar results.

## 2 The methodological framework

### 2.1 The Minimal Spanning Tree (MST)

The MST can be seen as a modern extension of the Horizontal- (or Vertical) Hierarchical-Tree-Plot – an older clustering method well known for its large applicability in medicine, psychiatry and archaeology (Hartigan, 1975). The essential additional ingredients of MST consist in the use of the ultrametric subdominant space and of the ultrametric distance between objects.

In order to clarify the role of the above ingredients, let's consider a system composed of $N$ agents (countries, regions, industrial branches, etc). Then, the classical MST can be constructed in the following steps:

(i) First, calculate the statistical distances $d_{ij}$ between all pair of agents (using e.g. Eq. 1.4, or other way of defining the statistical distance). Rank by increasing order the $N(N-1)/2$ values of the statistical distances $d_{ij}$.

(ii) Pick the pair corresponding to the smallest $d_{ij}$ and create a link between these two agents. Take the second smallest pair, and create a link between these two. Repeat the operation *unless* adding a link between the pair under consideration creates a loop in the graph, in which case one skips that value of $d_{ij}$. In other words, every new agent is added to the structure *only if* it has not been already included there.

(iii) Once all stocks have been linked at least once without creating loops, on finds a tree which only contains the strongest possible correlations, called the Minimum Spanning Tree. An example of this construction is shown in Fig. 1a.

Now, clusters can easily be created by removing all links of the MST such that $d_{ij} > d^*$. Since the tree contains no loops, removing one link cuts the tree into disconnected pieces. The remaining pieces are clusters within which all remaining links are "strong", i.e. such that $d_{ij} < d^*$ (or, equivalently, $C_{ij} > C^*$), which can be considered as strongly correlated. The number of disconnected pieces grows as the correlation threshold $d^*$ decreases.

Let us observe that the above structure is *not* Euclidean. In a Euclidean metrics the well known relations:

$$\begin{cases} d_{ij} = 0 \Leftrightarrow i = j; \\ d_{ij} = d_{ji}; \\ d_{ij} \leq d_{ik} + d_{kj}. \end{cases} \quad (6)$$

hold. However, in MST the last inequality ("the triangle inequality") is replaced by a stronger one, called "the ultrametric inequality", such that the above relations must be read:

$$\begin{cases} \hat{d}_{ij} = 0 \Leftrightarrow i = j; \\ \hat{d}_{ij} = \hat{d}_{ji}; \\ \hat{d}_{ij} \leq \max\{\hat{d}_{ik}, \hat{d}_{kj}\}. \end{cases} \quad (7)$$

The ultrametric spaces offer a natural description of the hierarchically structured complex systems as the concept of "ultrametricity" is directly related to the concept of "hierarchy"[2].

One first problem with the MST is that one often ends up with clusters of very dissimilar sizes. This aspect can lead either to a maximal dispersed structure (each object is in a class by itself) or, contrarily, to a high clustered structure in which all objects are joined together[3].

The MST was used in Hill (2001) as a methodology for linking countries together, so that international price and quantity indexes were chained. In Hill's approach the graph must not contain loops to ensure that the multilateral price indexes are transitive and hence internally consistent. The countries were grouped in two samples: the first consisted of 10 from Western Europe, 3 from Eastern Europe, 2 from North America, 7 from Asia and 8 from Africa; the second included the European countries and some former Soviet republics. The author concluded that "chaining can considerably simplify, and cut the cost of, multilateral international comparisons, while at the same time increasing characteristicity."

MST was also used in Andersen (2003) for linking together various industrial branches, with explicitly references to Darwinian phenograms and phylograms. The trees were (re-) constructed by means of input characteristics and output characteristics and then they are compared both with each other and with the industrial classification scheme (ISIC). One may be note here that, in general, biologists focus their interest more on the *shape* of the (phylogenetic) tree rather than on the *distance* between vertices of the tree because "it is more important in this context to assess the existence of common ancestors rather than to suggest when the separation of the species did occur" (Abdi, 1990). On the contrary, Andersen's approach offers a valuable suggestion of how to study the evolutionary transformation of the European industry.

One may also mention here the MST application in the stock market framework (Mantegna, 1999). Studying the MST and the hierarchical organization of the stocks defining the Dow Jones industrial average, Mantegna showed that the stocks can be divided into three groups. Carrying the same analysis for the stocks belonging to the S&P500, he obtained clusters of the stocks according to the industry they belong to.

**2.2 The robustness of MST and some complementary approaches**

Unlike the high frequency financial data series, the macroeconomic time series are too short and noisy. Most macroeconomic data have a yearly or at most quarterly frequency. A proper way for investigating such time series is by moving a constant size time window with a constant step so that the whole time interval is scanned.

The problem of MST robustness was explicitly addressed in Hill (2001). By comparing the MST for 1980 and 1985, and then for 1993 and 1996, the author concludes that "clearly the minimum spanning tree is not stable over time. Neither is it likely to be robust to slight changes in the data. This can be seen from Kruskal's algorithm. Any change in the ranking of the PLS$jk$ (Paasche-Laspeyres spread) measures may alter the minimum-spanning tree". This lack of robustness is also noticed in Andersen (2003) when the trees are compared over time and across countries. Here, the author uses the changes of the tree shape for drawing conclusions about the evolutionary process of (European) economic transformation.

---

[2] The connections between the ultrametric spaces and the indexed hierarchies were rigorous studied in Benzécri (1984).

[3] Nonetheless, the fact that clusters have dissimilar sizes may be a reality, related to the organization of the economic activity as a whole (Bouchaud and Potters, 2003).

In Figs. 1a-1b the MSTs[4] referring to the GDP data between 1994 and 2003 are shown. One can easily see that the shape of the trees strongly depends on the tree root choosing.

Some alternative ways for constructing the hierarchy, better adapted to the low frequency time series have been recently proposed. The Local Minimum Spanning Tree (LMST) is a modification of the MST algorithm under the constraint that the initial pair of nodes (the root) of the tree is the pair with the strongest correlation. Correlation chains have been investigated in the context of the most developed countries clustering in two forms: unidirectional and bidirectional minimum length chains (UMLP and BMLP respectively) (Miskiewicz and Ausloos, 2005). UMLP and BMLP algorithms are simplifications for LMST, where the closest neighbouring countries are attached at the end of a chain. In the case of the unidirectional chain the initial node is an arbitrary chosen country. Therefore in the case of UMLP the chain is expanded in one direction only, whereas in the bidirectional case countries might be attached at one of both ends depending on the distance value. These authors also underlined some arbitrariness in the root of the tree for comparing results, and considered that an a priori more common root, like the sum of the data, called the "All" country, from which to let the tree grow was permitting a better comparison.

## 2.3 The Moving-Average Minimal Length Path (MAMLP) Method

The problem that MST cannot be built in a unique way becomes even more important when we try to construct a cluster hierarchy for each position of a moving time window. The hierarchical structure proves to be not robust when the time window is moved even a single one year time step (see Figs. 1a and 1c). Simply, if the statistical distances between pairs A-B and C-D belonging to different clusters are small, it is quite likely to find at the next time step A-C and B-D as pairs in other different clusters.

In the MAMLP method described here below we propose to construct the hierarchy also starting from a virtual 'average' agent. In fact, the method of decoupling the movement of the weight centre of the system and the movement of independent parts is quite of common use in science.

The method is developed in the following steps:
(i) An 'AVERAGE' agent (AV) is virtually included into the system;
(ii) The statistical distance matrix is constructed, and thereafter, the elements are set into increasing order (i.e. the decreasing order of correlations);
(iii) The hierarchy is constructed, connecting each agent by its minimal length path to AV. Its minimal distance to AV is associated to each agent (see Fig.1d).
(iv) The procedure is repeated by moving a given and constant time window over the investigated time span.
(v) The agents are sorted through their movement inside the hierarchy. A new correlation matrix between country distances to their own mean is therefore constructed (see Subsection 3.3).

---

[4] The MSTs in Figs. 1a – 1c were constructed using MEGA soft (see the Andersen's project on the use of phylogenetic/ phenetic methods in evolutionary economics at:
http://www.business.aau.dk/evolution/projects/phylo/index.html)

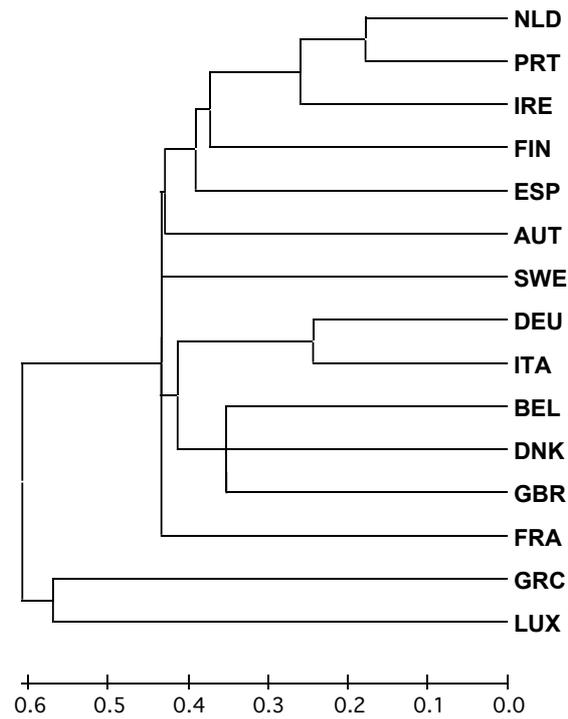

**Figure 1a** The MST of EU-15 countries for the time window 1994-2003. Indicator: GDP. The root of the branch is LUX

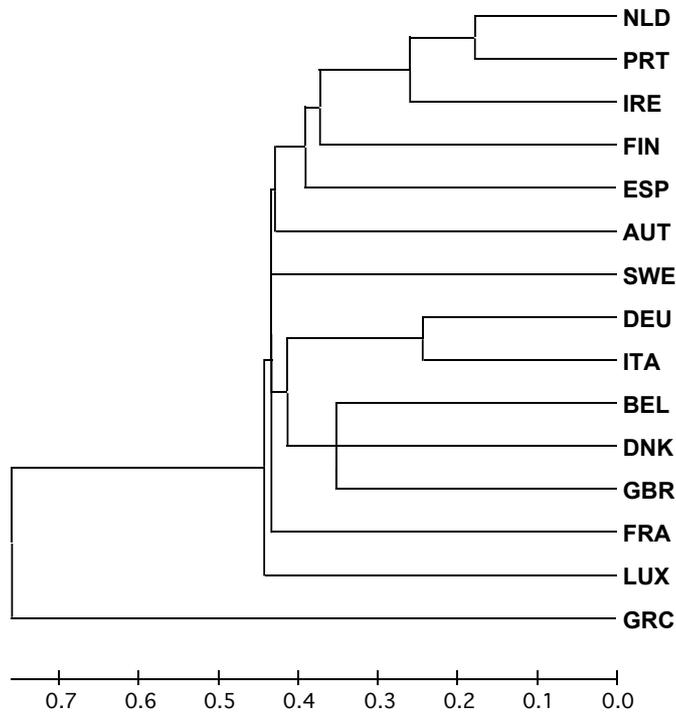

**Figure 1b** The MST of EU-15 countries for the time window 1994-2003. Indicator: GDP. The root on branch is GRC.

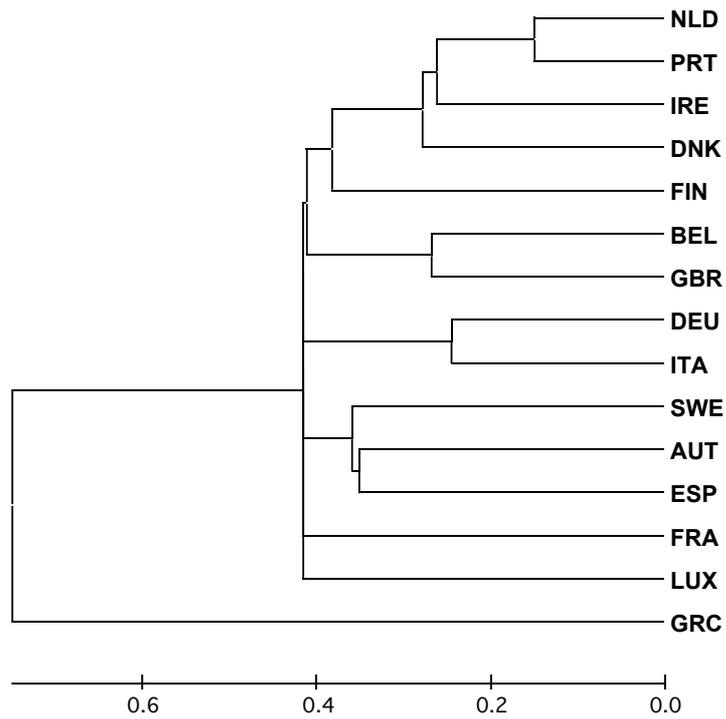

**Figure 1c** The MST of EU-15 countries for the time window 1995-2004. Indicator: GDP. The root of the branch is GRC.

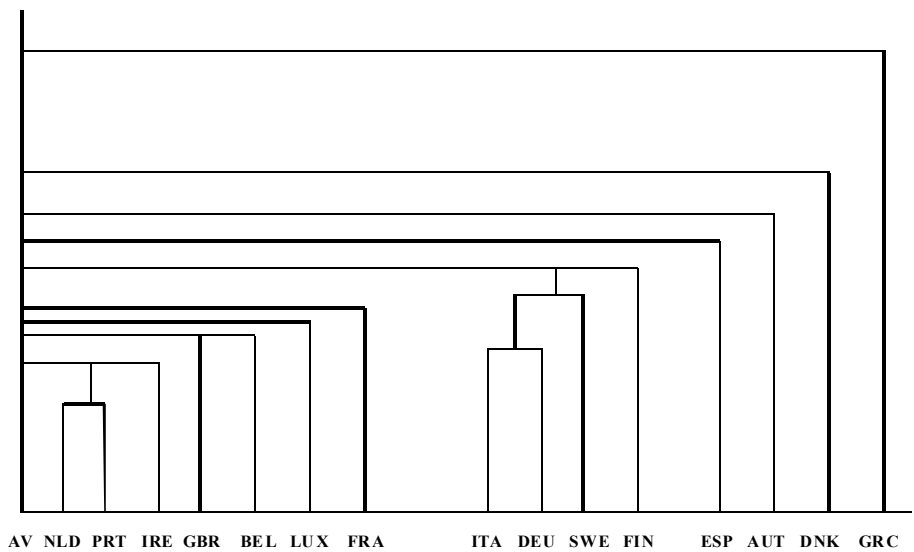

**Figure 1d** The MAMLP tree of EU-15 countries for the time window 1994-2003. Indicator: GDP.

# 3 Data processing and results

## 3.1 Data sources

The target group of countries is composed of 15 EU countries; the data refers to years between 1972 and 2004 (for the 10 years size time window analysis) and between 1994 and 2004 (for the 5 years size time window analysis case), that is before the last wave of EU extension.

The main source used for all the above indicators annual rates of growth taken between 1972 and 2004 is here below the World Bank database:
http://devdata.worldbank.org/query/default.htm.

In addition to the above mentioned data bank, for comparison aims, we also used the data supplied by:
http://www.economicswebinstitute.org/concepts.htm (1986-2000);
http://www.oecd.org/about/0,2337,en_2649_201185_1_1_1_1_1,00.html (2003-2004).

We abbreviate the countries according to The Roots Web Surname List (RSL) which uses 3 letters standardized abbreviations to designate countries and other regional locations (http://helpdesk.rootsweb.com/codes/). Inside the tables, for spacing reasons we use the countries two letters abbreviation (http://www.iso.org/iso/en/prods-services/iso3166ma/02iso-3166-code-lists/list-en1.html).

## 3.2 The mean statistical distance between EU countries in various time window sizes

GDP/capita data is first investigated with a fixed $T = 10$ years moving time window size, and the statistical distance matrix $D$ thereby constructed, taking into account $N = 15$ countries, namely AUT, BEL, DEU, DNK, ESP, FIN, FRA, GBR, GRC, IRL, ITA, LUX, NLD, PRT and SWE. The mean distance between the countries $<d>$ is calculated by averaging the statistical distances from $D$, over each time interval:

$$<d>_{(t,t+T)} = \frac{1}{N} \sum_{\substack{i,j=1 \\ i \neq j}}^{N} d_{ij} \qquad (8)$$

In order to identify the trend of $<d>$, we use the standardized mean statistical distance, defined as:

$$<\widetilde{d}>_{(t,t+T)} = \frac{1}{\sigma} <d>_{(t,t+T)} \qquad (9)$$

where:

$$\sigma \equiv \sigma(t,T) = \left[ \frac{1}{N} \sum_{\substack{i,j=1 \\ i>j}}^{N} [d_{ij} - <d>_{(t,t+T)}]^2 \right] \qquad (10)$$

is the standard deviation of the dataset.

In Figure 2 the standardized mean statistical distance is plotted taking into account all 15 EU-countries, between 1972 and 2004, by moving the 10 years time window by a one year time step. For simplicity, the interval notation is abbreviated at the last two digits of the first and last year of the window, and each data point is arbitrarily centred in the middle of the interval.

The time evolution of $<\tilde{d}>$ sets off a succession of abrupt increases ("shocks") followed by decreases ("relaxations"). Such phenomenon, occurred in the time interval 1986-2004, is separately plotted in Figure 3. The variable $x$ of the fit function (in the inset) represents the order number of the point. The time variation of $<\tilde{d}>$ displays an unexpected abrupt jump when going from 1991-2000 to 1992-2001, followed by a decay well fitted by an exponential (see inset). If the exponential decay is written as: $<\tilde{d}> = (const)\exp(-x/\tau)$, then $\tau$ is often called "the relaxation time" of the process. Here it is about 12.5 years. The abrupt jump of $<\tilde{d}>$ in Figure 3 between 91-00 and 92-02 occurs together with some similar anomaly in other statistical properties of the $\{d_{ij}\}$ datasets, as the variance, kurtosis and skewness (see Figure 4). Suspecting an effect due to Germany reunification, the data has been reanalyzed and is also shown on the same figure, but for only 14 countries (removing DEU – Figure 3), - but the anomalies remain.

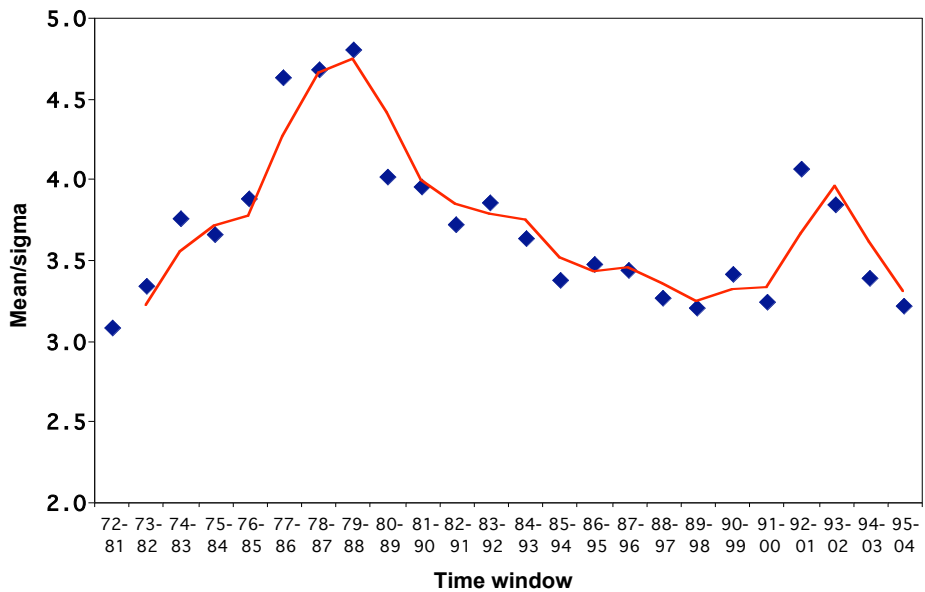

**Figure 2:** The GDP/capita standardized mean statistical distance of EU-15 countries from 1972 to 2004 corresponding to a 10 years moving time window. The line represents the 2-step mobile average fit.

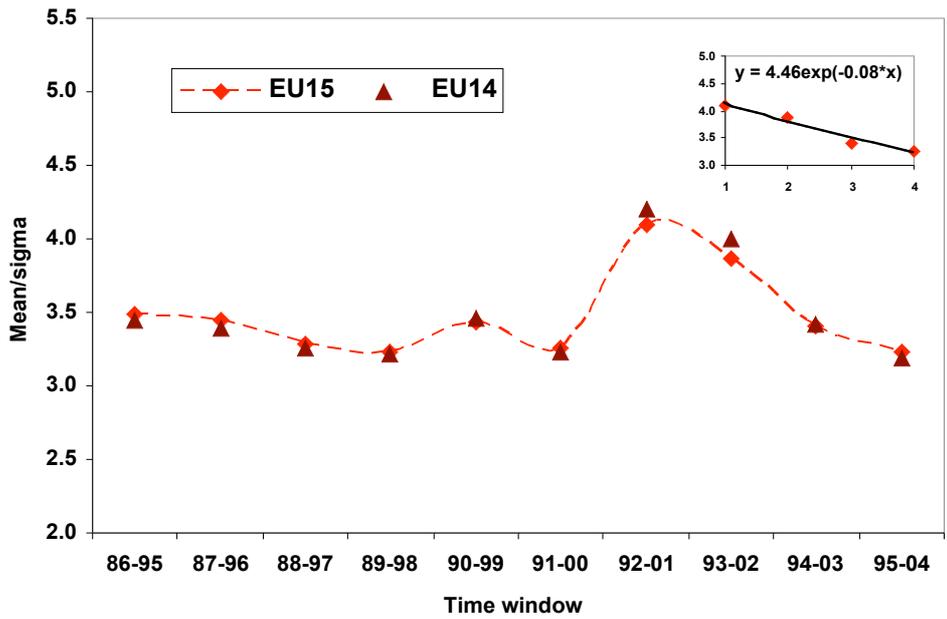

**Figure 3:** The GDP/capita standardized mean statistical distance of the EU-15 countries (diamond symbol) and EU-14 countries (triangle) respectively (removing DEU), from 1986 to 2004 corresponding to a 10 years moving time window. The inset represents the last 4 points of the main graph, fitted by an exponential. The Pearson RSQ fitting coefficient 0.97.

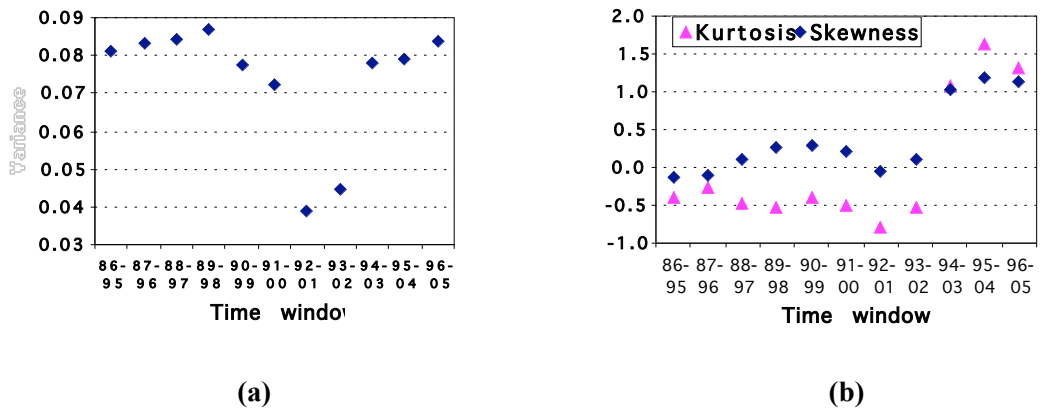

(a)  (b)

**Figure 4:** Evolution of the common characteristics (variance, kurtosis, skewness) of the distribution of statistical distances in the case of the GDP/capita of EU-15 countries, from 1986 to 2004, shown for a moving 10 years time window.

In the next step of investigation, the second branch, *i.e.* the time interval 1994-2004, is scanned with a shorter 5 years moving time window. A monotonic decreasing trend is again easily noticeable in Figure 5, corresponding to a relaxation time of the same order of magnitude, i.e., $\tau \sim 8\text{-}10$ years.

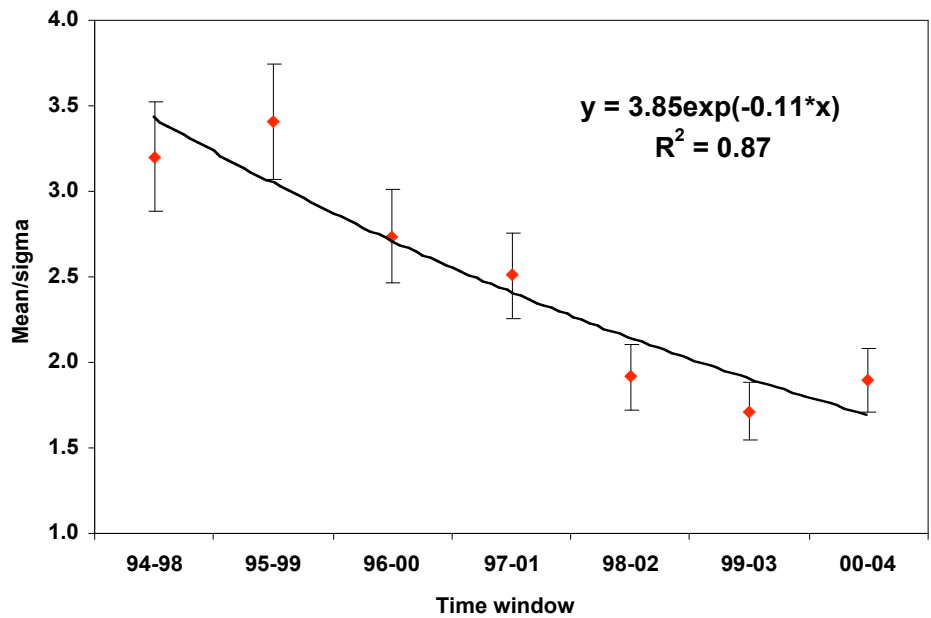

**Figure 5:** The GDP/capita standardized mean statistical distance of the EU-15 countries from 1994 to 2004 corresponding to a 5 years moving time window. The variable *x* of fit function is the order number of point. $R^2$ is the Pearson RSQ fitting coefficient. Error bars are bootstrap 90% confidence intervals.

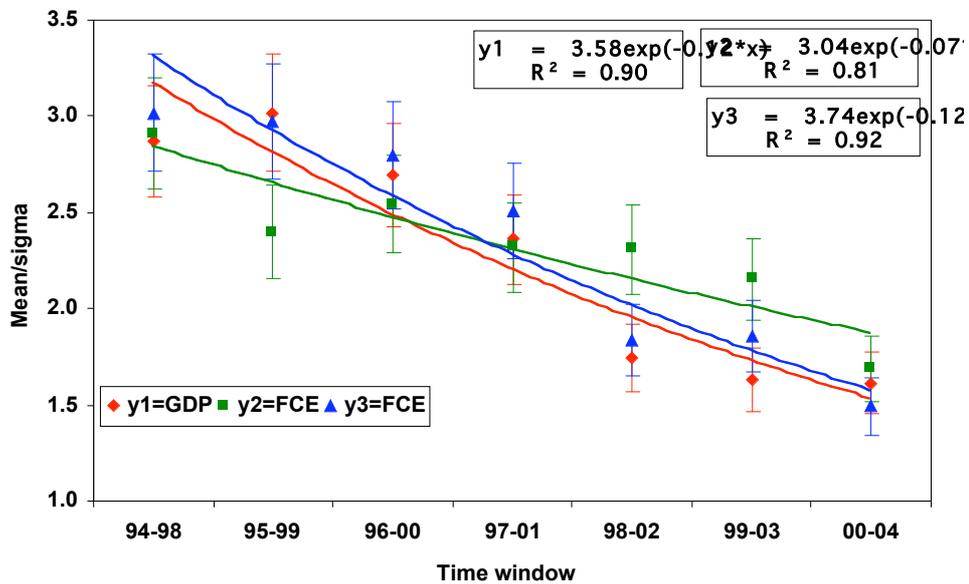

**Figure 6:** The GDP, FCE and GCF standardized mean statistical distance of the EU-15 countries from 1994 to 2004 corresponding to a 5 years moving time window. The variable *x* of the exponential fit function is the order number of point. $R^2$ is the Pearson RSQ coefficient of fitting. Error bars are bootstrap 90% confidence intervals.

In view of this time window effect, it seems reasonable to study the mean statistical distance between countries using GDP, CONS and GCF annual growth rates for the same (short) 5 years moving time window, for the data taken from 1994 to 2004[5].

It is seen that the standardized mean distance among the EU-15 countries, as plotted in Figure 6, follows the same decreasing trend as in Figure 5 for the GDP/capita, indicating a remarkable degree of similarity between the after-shock responses of the system with respect to GDP and GCF fluctuations (the same relaxation time $\tau \sim$ 8-10 years is found as in the case of GDP/capita). The relaxation time is $\tau > 10$ years for FCE fluctuations. We recall here that the term "fluctuations" refers, as above, to the annual rates of growth of the considered indicators (see data in insets).

Analyzing the time evolution of the mean statistical distance between the EU-15 countries one expects to find a decreasing trend, when one expects a global economic convergence. For the 10 years moving time window size (Figures 2 and 3) one can see a decreasing trend between 1979 and 1992 and for the last 4 time intervals, *i.e.*, the period 1992-2004, when the mean distance decreases from 4.80 to 3.20 and from 4.09 to 3.06 respectively (in $m/\sigma$ units, where $m$ = the mean and $\sigma$ = the standard deviation). In return, taking into account the whole evolution, the phenomenon appears as strongly nonlinear and non-monotonic. A somewhat unexpected evolution is registered in 1991-2000 and 1992-2001, when the mean distance abruptly increases (in a single step) from 3.26 to 4.09. It is not only a change of value but also a change of trend (Figure 3), i.e., from a quasi-constant trend (or a slow linear decrease) to another one that is strongly decreasing well fitted to an exponential. The abrupt change of trend also occurred for other statistical parameters of the distance distributions, *e.g.* the variance, kurtosis and skewness (Figure 4), approximately in the same time interval or in the next one.

The first explanation one could imagine would be the Berlin Wall fall and Germany re-unification. Indeed, Germany was taken into consideration in the previous estimation of the mean distance and by far, it was having the most abrupt variation of economic parameters in that period (see e.g. Keller, 1997). But the phenomenon seems to be somewhat more complex. In Figure 3 it has been seen that the time variation of the mean distance between countries with or without Germany (and its connections) (the EU-14 plot), is not at all affected. Another explanation might be found when analyzing several other important events which occurred *after* the Berlin wall fall *i.e.* the political changes and opening of new markets in Eastern Europe and Central Asia, while the Western European countries and their investors were having different positions in relation with these new possibilities of investment[6].

On the contrary, when a 5 years time window size is moved over the interval 1994-2004, there is a clear decrease of the mean statistical distance between EU-15 countries from 3.20 to 1.89 as concerns GDP/capita (Figure 5), from 2.86 to 1.81 for GDP, from 2.91 to 1.68 for the Final Consumption and from 3.01 to 1.49 for the Capital Growth (Figure 6). The mean distance does not display a clear trend as regards Net Exports fluctuations – at least in this time window size.

---

[5] In our used database, the Gross Capital Formation and the Net Exports data are available, for several of the considered countries, until 2003. Therefore, for these two indicators, the last time interval is taken from 2000 to 2003, i.e. for a 4 years time interval.

[6] This diffusion process generating an abrupt increase of the mean distance between countries was described in ACP model (Ausloos et al., 2004). It is interesting to note that in physical models these nonequilibrium abrupt transitions, due to "shocks", are generally followed by exponential or power law relaxations, (Lambiotte and Ausloos, 2006; Sornette et al., 2004).

## 3.3 Country clustering structure along the MAMLP method

At this point of our investigation the subsequent ingredients of the MAMLP method, introduced in Sect. 2, are implemented. The first indicator taken into consideration is the GDP annual growth. A virtual 'AVERAGE' country is introduced in the system. The statistical distances corresponding to the fixed 5 years moving time window are set in increasing order and the minimal length path (MPL) connections to the AVERAGE are established for each country in every time interval (Table 1).

Table 1: MLP distances to AVERAGE. Indicator: GDP. The moving time window size is 5 years for data taken from 1994 to 2004.

|       | AT  | BE  | DE  | DK  | ES  | FI  | FR  | UK  | GR   | IE  | IT  | LU  | NL  | PT  | SE  |
|-------|-----|-----|-----|-----|-----|-----|-----|-----|------|-----|-----|-----|-----|-----|-----|
| 94-98 | .67 | .86 | .86 | .86 | .40 | .40 | .67 | .86 | .40  | .86 | .86 | .40 | .40 | .86 | .86 |
| 95-99 | .60 | .65 | .52 | .71 | .21 | .77 | .45 | .77 | .37  | .65 | .90 | .37 | .23 | .83 | .52 |
| 96-00 | .58 | .32 | .46 | .61 | .34 | .81 | .46 | .32 | .32  | .53 | .32 | .20 | .60 | .60 | .46 |
| 97-01 | .48 | .30 | .48 | .30 | .28 | .42 | .48 | .44 | .68  | .38 | .68 | .14 | .28 | .28 | .48 |
| 98-02 | .43 | .26 | .19 | .19 | .21 | .43 | .19 | .19 | 1.04 | .29 | .44 | .12 | .21 | .21 | .29 |
| 99-03 | .25 | .23 | .19 | .19 | .29 | .26 | .19 | .37 | 1.15 | .26 | .37 | .23 | .19 | .19 | .28 |
| 00-04 | .27 | .27 | .17 | .26 | .28 | .27 | .21 | .27 | .53  | .50 | .28 | .27 | .21 | .21 | .27 |

As one can see in Table 1, if the countries are ordered after the distances to AVERAGE, the resulting hierarchy is found to be changing from a time interval to another. Therefore, another correlation matrix is built, this time for the country movements inside the hierarchy. The matrix elements are defined as:

$$\hat{C}_{ij}(t) = \frac{<\hat{d}_i(t)\hat{d}_j(t)> - <\hat{d}_i(t)><\hat{d}_j(t)>}{\sqrt{<[\hat{d}_i(t)]^2 - <\hat{d}_i(t)>^2><[\hat{d}_j(t)]^2 - <\hat{d}_j(t)>^2>}} \quad (11)$$

where $\hat{d}_i(t)$ and $\hat{d}_j(t)$ are the minimal length path (MPL) distances to the AVERAGE. For simplicity, in Eq. (11) are not included the explicit dependencies on the time window size $T$.

In this way the strongest correlations and anti-correlations between GDP fluctuations could be extracted and a clustering structure searched for.

Regarding the country clusters, as in other classification problems, a major issue that arises when the classification trees derive from real data with much random noise concerns how to define what a cluster is. This general issue is discussed in the literature on tree classification under the topic of *over-fitting* (Breiman et al., 1984) If not stopped, the tree algorithm will ultimately "extract" all information from the data, including random or noise variation.

To avoid this trap in our classification we have considered as "strong" correlations and anti-correlations those with $C \geq 0.9$ and $C \leq -0.5$ respectively, taking into account that the both intervals of $C$ include the same percentage (~ 10 %) from the total set of correlation coefficients. From this criterion, the strongly correlated countries in GDP fluctuations (as indicated in bold faces in Table 2) can be partitioned into two clusters: FRA-SWE-DEU and BEL-GBR-IRE-DNK-PRT. ITA can be considered in the second cluster for its strong correlation with GBR, but it does not display any strong correlations with the other countries. LUX is weakly correlated to the second cluster, while AUT is somewhat "equidistant" displaying medium correlations with both clusters. GRC holds a special position: its GDP fluctuations appear to be strongly anti-correlated with of all other countries.

Table 2: The correlation matrix of country movements inside the hierarchy; Indicator: GDP. The moving time window size is 5 years for data taken from 1994 to 2004.

|    | AT | BE  | DE  | DK  | ES  | FI   | FR  | UK  | GR   | IE   | IT   | LU   | NL   | PT   | SE   |
|----|----|-----|-----|-----|-----|------|-----|-----|------|------|------|------|------|------|------|
| AT | 1  | .77 | .88 | .88 | .33 | .69  | .88 | .69 | -.69 | .75  | .71  | .42  | .61  | .89  | .85  |
| BE |    | 1   | .88 | **.90** | .41 | .27 | .80 | **.94** | -.59 | **.92** | .83 | .85 | .23 | **.90** | **.91** |
| DE |    |     | 1   | **.90** | .61 | .35 | **.98** | .86 | -.65 | .85 | .78 | .61 | .52 | .86 | **.99** |
| DK |    |     |     | 1   | .50 | .58  | .87 | .84 | -.80 | **.93** | .67 | .77 | .58 | **.99** | .88 |
| ES |    |     |     |     | 1   | -.10 | .61 | .34 | -.38 | .55  | .05  | .36  | .66  | .37  | .64  |
| FI |    |     |     |     |     | 1    | .42 | .25 | **-.62** | .34 | .27 | .14 | .60 | .64 | .26 |
| FR |    |     |     |     |     |      | 1   | .79 | **-.71** | .81 | .73 | .52 | .60 | .82 | **.95** |
| UK |    |     |     |     |     |      |     | 1   | **-.52** | .82 | **.90** | .85 | .12 | .86 | .86 |
| GR |    |     |     |     |     |      |     |     | 1    | **-.82** | -.38 | **-.56** | **-.62** | **-.76** | **-.60** |
| IE |    |     |     |     |     |      |     |     |      | 1    | .63  | .85  | .43  | .89  | .87  |
| IT |    |     |     |     |     |      |     |     |      |      | 1    | .59  | -.05 | .73  | .77  |
| LU |    |     |     |     |     |      |     |     |      |      |      | 1    | .06  | .77  | .65  |
| NL |    |     |     |     |     |      |     |     |      |      |      |      | 1.   | .50  | .47  |
| PT |    |     |     |     |     |      |     |     |      |      |      |      |      | 1    | .84  |
| SE |    |     |     |     |     |      |     |     |      |      |      |      |      |      | 1    |

The MAMLP method can now be applied to the other three macroeconomic indicators defined in Section 2, namely Final Consumption Expenditure, Gross Capital Formation and Net Exports. Tables A2, A4 and A6 give the corresponding MLP distances to AVERAGE, while Tables A3, A5 and A7 display the correlation matrices. As for Table 2, Tables A3, A5 and A7 display in bold the strongest correlations and anticorrelations.

In the above mentioned tables we can observe the position of the bold elements, whence see that five of the mostly correlated countries with respect to GDP fluctuations (SWE-GBR-DEU-BEL-IRL) also display strong correlations in the Final Consumption Expenditure and medium correlations in Gross Capital Formation fluctuations ($C_{ij} \sim 0.8$). Moreover, some of them are strongly anticorrelated in Net Exports fluctuations (e.g. $C_{ij} < -0.9$ for DEU-SWE and DEU-IRL). The top strong correlations appear in FCE fluctuations (Table A3), while the top anticorrelations can be found in NEX fluctuations (Table A7).

Finally, we calculate aso called *sensitivity degree*, i.e., the quadratic sum of all the correlation coefficients:

$$(\chi_i)_\alpha = \sum_{\substack{i,j=1 \\ i \neq j}}^{N} (\hat{C}_{ij})^2 \qquad (12)$$

where $\alpha \equiv$ GDP, FCE, GCF and NEX. The results are given in Table 3 for all considered indicators and for each country.

Table 3: The quadratic sum of correlation coefficients (the *sensitivity degree* of countries) for the fluctuations of GDP, Final Consumption Expenditure (FCE), Gross Capital Formation

(GCF) and Net Exports (NEX), for data taken from 1994 to 2004 (GDP and FCE) and from 1994 to 2003 (GCF and NEX) respectively..

| | GDP | | FCE | | GCF | | NEX |
|---|---|---|---|---|---|---|---|
| DK | 9.08 | BE | 8.34 | AT | 4.99 | PT | 5.23 |
| PT | 8.71 | IE | 8.34 | SE | 4.69 | DE | 4.92 |
| DE | 8.68 | ES | 8.32 | ES | 4.66 | IE | 4.76 |
| SE | 8.47 | NL | 8.32 | FR | 4.66 | SE | 4.76 |
| IE | 8.26 | PT | 8.32 | BE | 4.58 | IT | 4.41 |
| BE | 8.25 | SE | 8.32 | DK | 4.18 | AT | 3.99 |
| FR | 8.21 | UK | 8.14 | FI | 4.09 | DK | 3.50 |
| AT | 7.60 | DE | 7.42 | IE | 3.04 | FR | 3.24 |
| UK | 7.59 | AT | 7.15 | PT | 2.89 | FI | 3.23 |
| IT | 5.68 | FR | 3.07 | DE | 2.85 | LU | 3.23 |
| GR | 5.64 | FI | 3.06 | IT | 2.70 | UK | 2.91 |
| LU | 5.40 | LU | 1.81 | UK | 2.68 | BE | 2.71 |
| NL | 3.25 | DK | 1.61 | GR | 2.63 | NL | 2.63 |
| ES | 2.97 | GR | 1.60 | LU | 2.39 | GR | 2.49 |
| FI | 2.68 | IT | 1.13 | NL | 2.31 | ES | 1.69 |

One can note that the sensitivity classifications regarding GDP and Net Exports fluctuations are quite similar, at least for the countries situated at the top and at the bottom. We recover here one of the main characteristics of social networks that is the positive correlation existing between the node degrees (Ramasco et al., 2003), *i.e.* the highly connected countries commonly tend to connect with other well connected ones. So, a new empirical evidence of regional convergence clubs is hereby found.

# 4 Conclusion

In the present study, the mean statistical distance between countries was defined on the support of their macroeconomic fluctuations and a new statistical methodology, called MAMPL method, was applied. We can resume our findings as follows:
(1) The decreasing of the mean statistical distance between EU countries is reflected in the correlated fluctuations of the basic ME indicators: GDP, GDP/capita, Consumption and Investments; this empirical evidence can be seen as an economic aspect of globalization.
(2) The increasing and decreasing of the mean statistical distance between EU countries occur cyclically, being strongly influenced by the economic booms and busts as well as by endogenous and exogenous shocks (induced by the political and institutional shifts)
(3) Even inside of the apparently homogeneous region of development, (*e.g.* the Western Europe), a spontaneous country clustering occurs.
   The choosing of the macroeconomic variables is motivated by the fact the economic performance of any country is most frequently evaluated in the terms of GDP, investments, consumption and trade. As well as many economists, sociologists, politicians, etc. have already done, we may wonder: is the globalization a real phenomenon or it is only an analytical artefact (a myth)? Our premise is that if there is a *real* convergence of countries, it must be somehow embodied in the time evolution of the basic macroeconomic indicators. If this is the case, a new problem here arises, related to the optimal way of extracting the information from the sparse and noisy macroeconomic time series. The question of the optimal choosing of the time window size, as well as the one of deriving an adequate methodology for constructing the country classification tree, are explicitly broached in the text of this paper.

Also, as long as we consider only time variation of the macroeconomic indicators, without taking into account the regional factors (e.g. the geographic distances), a theoretical approach can remain essentially at the one-dimensional level of description. In the present approach, the ME time series are seen as outputs embodying all manner of interactions between countries (e.g. the technology, R&D and information spill-over among countries or regions). This kind of (descriptive) approach does not allow for introducing control variables, whence political and institutional shifts induced by EMU are not explicitly accounted for. Further developments of the present approach will have to consider both spatial and dynamic correlations jointly on the line suggested in Roehner (1993) and Quah (1996). A way for taking into account multivariate analysis framework may also be the bi-partite factor graph described in Appendix 3.

Beyond the novelties in the cluster analysis methodology, there are several additional policy implications of our empirical findings, which we wish to highlight and discuss.

Firstly, the economic clusters arise in the presence of Marshallian externalities that signify that firms benefit from the production and innovation activities of neighboring firms in the same and related industries. There is a strong interaction between growth and clustering. For example, agglomeration and growth are mutually self-reinforcing, so that trade (with transportation costs) may lead to both higher growth and agglomeration. As the recent evolution of the developed countries has shown, instead of policies to reallocate resources across sectors, a better way is to implement policies to promote clustering in sectors that already show comparative advantage. This implies that, as generally accepted by proponents of cluster-based policies, governments should not try to create clusters starting from scratch.

On the same idea, promoting a cluster is not necessarily welfare enhancing, since it could be a cluster without a comparative advantage. When there are comparative advantages coming from sources different than clustering, promoting the creation of a cluster by distorting the prices so as to push resources into advanced sectors may be inferior to the status quo, and is always dominated by promotion of a cluster in sectors where the economy is already showing comparative advantage.

Trade shares, export shares, and import shares in GDP are widely used in the literature and are significantly and positively correlated with growth. There is also a positive and strong relationship between trade barriers and growth. One of the possible explanations is that if tariffs cause a reallocation of productive resources to the goods in which a country has comparative advantage from the goods in which a country has no advantage, then tariffs are likely to affect growth positively. This result also provides support for the infant industry case for protection and for strategic trade policies.

Recent research suggests that there are significant external sources of growth, which extend beyond borders. In particular, regional external economies from both physical and human capital accumulation are important for explaining differences in growth rates across countries. Since uncompensated spillovers play an important role in the process of economic development, economic integration can be an important driving force for growth. A cluster-promotion policy includes R&D incentives in the form of tax breaks and matching grants for both individual and collaborative innovation projects. A more ambitious policy would encourage and partially finance a long-term strategy for research and the creation of skills between the relevant industry associations and the most important universities and research centers.

# APPENDIX 1
## Shuffled data analysis

For a robustness test and statistical error bar significance, the elements of the statistical distance matrices were shuffled per columns so as the data proceeded from different time windows were randomly mixed. In all three index cases so considered, the mean distance derived from the shuffled data midly oscillates around a constant value, as it has to be expected; the amplitude of the fluctuations is 0.49 units mean/sigma for GDP, 0.12 units for FCE and 0.28 units for GCF, that means 35 %, 9.7 %, and 21.5 % respectively from their maximal (real) variation induced by the decreasing trend.

As a second test, the correlation matrix from Table 2 was randomized by shuffling MLP distances to AVERAGE (from Table 1), firstly per columns and secondly per lines. The results are presented in Table A1. The maximum and minimum values of the correlation coefficients are found to be $(C_{max})_{shufll} = 0.71$ and $(C_{min})_{shufll} = -0.68$ as compared with $(C_{max}) = 0.99$ and $(C_{min}) = -0.80$ from Table 2. According to the criterion discussed in Section 3 ($C_{corr} \geq 0.9$ and $C_{anticorr} \leq -0.8$), one can say that neither any strong correlations nor anti-correlation appear. In other words, the correlations which resulted in the clustering structure discussed in Section 3 are destroyed by the randomization, consequently giving weight to the main text results, analysis and conclusion.

Table A1: The randomized correlation matrix of country movements of inside the hierarchy. Indicator: GDP. Time window size: 5 years

|    | AT | BE  | DE   | DK   | ES   | FI   | FR   | UK   | GR   | IE   | IT   | LU   | NL   | PT   | SE   |
|----|----|-----|------|------|------|------|------|------|------|------|------|------|------|------|------|
| AT | 1  | .19 | -.07 | -.28 | .23  | -.23 | .45  | .55  | -.47 | .07  | -.35 | .28  | -.43 | .29  | -.49 |
| BE |    | 1   | .51  | .10  | -.10 | -.47 | .16  | .24  | -.35 | -.48 | -.61 | .41  | .07  | -.55 | .18  |
| DE |    |     | 1    | .53  | .24  | -.22 | **.70** | -.22 | -.48 | -.50 | -.11 | -.34 | -.02 | .24  | .16  |
| DK |    |     |      | 1    | -.32 | .19  | .19  | .27  | -.20 | -.64 | -.22 | -.67 | -.15 | .36  | .34  |
| ES |    |     |      |      | 1    | .42  | .58  | -.57 | -.60 | .32  | .66  | -.21 | .06  | .37  | .15  |
| FI |    |     |      |      |      | 1    | .00  | -.16 | -.17 | -.02 | **.71** | -.67 | .28  | .33  | .43  |
| FR |    |     |      |      |      |      | 1    | -.06 | -.53 | -.33 | .17  | -.44 | .00  | .62  | -.32 |
| UK |    |     |      |      |      |      |      | 1    | .00  | -.46 | **-.68** | .09  | -.23 | .00  | -.32 |
| GR |    |     |      |      |      |      |      |      | 1    | -.05 | .08  | .10  | .50  | -.37 | -.42 |
| IE |    |     |      |      |      |      |      |      |      | 1    | .26  | .44  | -.44 | .05  | .08  |
| IT |    |     |      |      |      |      |      |      |      |      | 1    | -.52 | .47  | .32  | .10  |
| LU |    |     |      |      |      |      |      |      |      |      |      | 1    | -.22 | **-.67** | -.12 |
| NL |    |     |      |      |      |      |      |      |      |      |      |      | 1    | -.40 | -.12 |
| PT |    |     |      |      |      |      |      |      |      |      |      |      |      | 1    | -.21 |
| SE |    |     |      |      |      |      |      |      |      |      |      |      |      |      | 1    |

# APPENDIX 2
## The MAMLP distances to AVERAGE and the correlation matrices for FCE, GCF, and NEX

Table A2: MLP distances to AVERAGE. Indicator: Final Consumption Expenditure. The moving time window size is 5 years for data taken from 1994 to 2004.

|       | AT   | BE   | DE   | DK   | ES   | FI   | FR   | UK   | GR   | IE   | IT   | LU   | NL   | PT   | SE   |
|-------|------|------|------|------|------|------|------|------|------|------|------|------|------|------|------|
| **94-98** | .88  | .65  | .85  | .88  | .65  | .37  | .65  | .65  | .65  | .65  | .37  | .65  | .65  | .65  | .65  |
| **95-99** | .79  | .79  | .79  | .81  | .79  | .41  | .79  | .79  | .93  | .79  | .53  | .59  | .79  | .79  | .79  |
| **96-00** | 1.02 | 1.02 | 1.02 | 1.02 | 1.02 | 1.02 | 1.02 | 1.02 | 1.02 | 1.02 | .26  | 1.02 | 1.02 | 1.02 | 1.02 |
| **97-01** | .51  | .51  | .51  | .65  | .51  | .73  | .88  | .51  | .65  | .51  | .33  | .88  | .51  | .51  | .51  |
| **98-02** | .52  | .52  | .52  | .96  | .52  | .66  | .95  | .65  | .96  | .52  | .35  | 1.19 | .52  | .52  | .52  |
| **99-03** | .45  | .42  | .45  | 1.00 | .45  | .53  | .40  | .46  | 1.00 | .42  | .30  | .92  | .45  | .45  | .45  |
| **00-04** | .88  | .65  | .85  | .88  | .65  | .37  | .65  | .65  | .65  | .65  | .37  | .65  | .65  | .65  | .65  |

Table A3: The correlation matrix of country movements inside the hierarchy. Indicator: Final Consumption Expenditure. The moving time window size is 5 years for data taken from 1994 to 2004.

|     | AT | BE  | DE  | DK  | ES  | FI  | FR  | UK  | GR  | IE  | IT  | LU   | NL  | PT  | SE  |
|-----|----|-----|-----|-----|-----|-----|-----|-----|-----|-----|-----|------|-----|-----|-----|
| AT  | 1  | .92 | 1   | .23 | .92 | .21 | .38 | .87 | .03 | .92 | .07 | -.34 | .92 | .92 | .92 |
| BE  |    | 1   | .94 | .23 | 1   | .45 | .56 | .97 | .28 | 1   | .06 | -.15 | 1   | 1   | 1   |
| DE  |    |     | 1   | .24 | .93 | .24 | .40 | .89 | .07 | .94 | .07 | -.32 | .93 | .93 | .93 |
| DK  |    |     |     | 1   | .26 | .22 | -.14| .35 | .75 | .23 | -.41| .44  | .26 | .26 | .26 |
| ES  |    |     |     |     | 1   | .45 | .53 | .97 | .31 | 1   | .04 | -.15 | 1   | 1   | 1   |
| FI  |    |     |     |     |     | 1   | .65 | .49 | .34 | .45 | **-.68** | .68 | .45 | .45 | .45 |
| FR  |    |     |     |     |     |     | 1   | .64 | .05 | .56 | -.05| .38  | .53 | .53 | .53 |
| UK  |    |     |     |     |     |     |     | 1   | .40 | .97 | .03 | .02  | .97 | .97 | .97 |
| GR  |    |     |     |     |     |     |     |     | 1   | .28 | -.11| .45  | .31 | .31 | .31 |
| IE  |    |     |     |     |     |     |     |     |     | 1   | .06 | -.15 | 1   | 1   | 1   |
| IT  |    |     |     |     |     |     |     |     |     |     | 1   | **-.68** | .04 | .04 | .04 |
| LU  |    |     |     |     |     |     |     |     |     |     |     | 1    | -.15| -.15| -.15|
| NL  |    |     |     |     |     |     |     |     |     |     |     |      | 1   | 1   | 1   |
| PT  |    |     |     |     |     |     |     |     |     |     |     |      |     | 1   | 1   |
| SE  |    |     |     |     |     |     |     |     |     |     |     |      |     |     | 1   |

Table A4: MLP distances to AVERAGE. Indicator: Gross Capital Formation. The moving time window size is 5 years for data taken from 1994 to 2003.

|       | AT  | BE  | DE  | DK  | ES  | FI  | FR  | UK  | GR  | IE  | IT  | LU  | NL  | PT  | SE  |
|-------|-----|-----|-----|-----|-----|-----|-----|-----|-----|-----|-----|-----|-----|-----|-----|
| **94-98** | .51 | .48 | .59 | .52 | .66 | .48 | .66 | .58 | .89 | .67 | .38 | .85 | .67 | .37 | .51 |
| **95-99** | .47 | .46 | .75 | .49 | .54 | .46 | .54 | .61 | .75 | .49 | .33 | .83 | .49 | .39 | .58 |
| **96-00** | .75 | .78 | .75 | .78 | .75 | .78 | .75 | .58 | .75 | .84 | .32 | .32 | .48 | .20 | .75 |
| **97-01** | .70 | .47 | .70 | .62 | .70 | .62 | .70 | .57 | .70 | .38 | .63 | .29 | .29 | .09 | .70 |
| **98-02** | .46 | .46 | .46 | .68 | .46 | .68 | .46 | .61 | .46 | .46 | 1.13| .46 | .46 | .46 | .46 |
| **99-03** | .70 | .70 | .70 | .88 | .70 | .88 | .70 | .70 | .70 | .70 | 1.07| .70 | .70 | .70 | .70 |

Table A5: The correlation matrix of country movements inside the hierarchy. Indicator: Gross Capital Formation. The moving time window size is 5 years for data taken from 1994 to 2003.

|     | AT | BE  | DE  | DK  | ES  | FI  | FR  | UK  | GR  | IE  | IT  | LU   | NL   | PT   | SE  |
|-----|----|-----|-----|-----|-----|-----|-----|-----|-----|-----|-----|------|------|------|-----|
| AT  | 1  | .76 | .59 | .68 | .88 | .69 | .88 | .10 | .19 | .45 | -.04| -.58 | -.12 | -.26 | **.94** |
| BE  |    | 1   | .47 | .81 | .67 | .79 | .67 | .35 | .15 | .85 | -.02| -.27 | .32  | .15  | .73 |
| DE  |    |     | 1   | .10 | .64 | .09 | .64 | .05 | .55 | .30 | -.57| -.02 | -.08 | -.25 | .81 |
| DK  |    |     |     | 1   | .41 | 1   | .41 | .61 | -.32| .50 | .56 | -.40 | .24  | .39  | .55 |
| ES  |    |     |     |     | 1   | .40 | 1   | -.04| .61 | .58 | -.35| -.26 | .11  | -.29 | .83 |

|    | AT | BE | DE | DK | ES | FI | FR | UK | GR | IE | IT | LU | NL | PT | SE |
|----|----|----|----|----|----|----|----|----|----|----|----|----|----|----|----|
| FI |    |    |    |    |    | 1  | .40 | .58 | -.37 | .46 | .57 | -.46 | .17 | .35 | .56 |
| FR |    |    |    |    |    |    | 1  | -.04 | .61 | .58 | -.35 | -.26 | .11 | -.29 | .83 |
| UK |    |    |    |    |    |    |    | 1  | -.21 | .20 | .63 | .37 | .61 | **.91** | .12 |
| GR |    |    |    |    |    |    |    |    | 1  | .44 | -.76 | .45 | .37 | -.20 | .27 |
| IE |    |    |    |    |    |    |    |    |    | 1  | -.26 | .10 | .62 | .21 | .40 |
| IT |    |    |    |    |    |    |    |    |    |    | 1  | -.15 | .12 | .60 | -.21 |
| LU |    |    |    |    |    |    |    |    |    |    |    | 1  | .73 | .60 | -.46 |
| NL |    |    |    |    |    |    |    |    |    |    |    |    | 1  | .78 | -.17 |
| PT |    |    |    |    |    |    |    |    |    |    |    |    |    | 1  | -.27 |
| SE |    |    |    |    |    |    |    |    |    |    |    |    |    |    | 1  |

Table A6: MLP distances to AVERAGE. Indicator: Net Exports. The moving time window size is 5 years for data taken from 1994 to 2003.

|       | AT   | BE  | DE  | DK   | ES  | FI  | FR  | UK  | GR  | IE  | IT  | LU  | NL   | PT  | SE  |
|-------|------|-----|-----|------|-----|-----|-----|-----|-----|-----|-----|-----|------|-----|-----|
| 94-98 | 1.27 | .19 | .65 | .89  | .45 | .80 | .65 | .62 | .75 | .62 | .62 | .80 | .64  | .62 | .62 |
| 95-99 | 1.13 | .40 | .66 | 1.11 | .66 | .87 | .66 | .56 | .87 | .56 | .56 | .87 | 1.11 | .56 | .56 |
| 96-00 | 1.29 | .72 | .52 | .81  | .52 | .81 | .56 | .22 | .81 | .72 | .54 | .81 | .54  | .54 | .72 |
| 97-01 | 1.06 | .55 | .64 | .80  | .64 | .70 | .64 | .26 | .39 | .55 | .64 | .70 | .64  | .64 | .55 |
| 98-02 | .94  | .73 | .54 | .73  | .54 | .67 | .73 | .54 | .54 | .73 | .54 | .67 | .67  | .54 | .73 |
| 99-03 | .37  | .65 | .37 | 1.03 | .50 | .82 | .79 | .76 | .65 | .79 | .50 | .82 | .82  | .37 | .79 |

Table A7: The correlation matrix of country movements inside the hierarchy. Indicator: Net Exports. The time moving window size is 5 years for data taken from 1994 to 2003.

|    | AT | BE | DE | DK | ES | FI | FR | UK | GR | IE | IT | LU | NL | PT | SE |
|----|----|----|----|----|----|----|----|----|----|----|----|----|----|----|----|
| AT | 1 | -.39 | .80 | -.32 | .11 | .02 | **-.89** | -.62 | .30 | **-.59** | .60 | .02 | -.26 | .84 | **-.59** |
| BE |   | 1 | **-.65** | -.39 | .09 | -.39 | .15 | -.30 | -.32 | .62 | **-.61** | -.39 | -.27 | -.48 | .62 |
| DE |   |   | 1 | -.07 | .44 | -.05 | -.56 | -.35 | .06 | -.92 | .82 | -.05 | .13 | **.93** | **-.92** |
| DK |   |   |   | 1 | .22 | .85 | .28 | .56 | .58 | -.14 | -.28 | .85 | .86 | -.41 | -.14 |
| ES |   |   |   |   | 1 | -.03 | -.16 | -.37 | -.18 | **-.64** | .23 | -.03 | .53 | .30 | **-.64** |
| FI |   |   |   |   |   | 1 | -.13 | .30 | .86 | -.04 | -.29 | **1** | .56 | -.31 | -.04 |
| FR |   |   |   |   |   |   | 1 | .82 | -.29 | .47 | -.47 | -.13 | .35 | **-.67** | .47 |
| UK |   |   |   |   |   |   |   | 1 | .21 | .34 | -.40 | .30 | .50 | **-.57** | .34 |
| GR |   |   |   |   |   |   |   |   | 1 | .05 | -.35 | .86 | .40 | -.16 | .05 |
| IE |   |   |   |   |   |   |   |   |   | 1 | **-.82** | -.04 | -.28 | **-.81** | **1** |
| IT |   |   |   |   |   |   |   |   |   |   | 1 | -.29 | -.24 | **.90** | **-.82** |
| LU |   |   |   |   |   |   |   |   |   |   |   | 1 | .56 | -.31 | -.04 |
| NL |   |   |   |   |   |   |   |   |   |   |   |   | 1 | -.25 | -.28 |
| PT |   |   |   |   |   |   |   |   |   |   |   |   |   | 1 | **-.81** |
| SE |   |   |   |   |   |   |   |   |   |   |   |   |   |   | 1 |

# APPENDIX 3
## Towards a multivariable approach. The Cluster Variation method

Let's consider a system with discrete degrees of freedom which will be denoted by $s = \{s_1, s_2,\ldots, s_N\}$. For instance, variables $s_i$ could take values in the set $\{0, 1\}$ (binary variables), $\{-1, +1\}$, or $\{1, 2, \ldots q\}$, $q \in N$.

The combinatorial optimization models are usually defined through a cost function $H = H(s)$, and the corresponding probability distribution is:

$$p(s) = \frac{1}{Z}\exp[-H(s)] \qquad (A1)$$

where:

$$Z = \sum_s \exp[-H(s)] \qquad (A2)$$

is the partition function.

The cost function (CF) is typically a sum of terms, each involving a small number of variables. A useful representation is given by the *factor graph*[7]. A factor graph is a bipartite graph (Lambiotte and Ausloos, 2005) made of variable nodes $i, j, \ldots$ one for each variable, and function nodes $a, b, \ldots$, one for each term of the cost function. In present approach the variable nodes are the macroeconomic indicators and the function nodes are the countries (Figure 7).

An edge[8] joins a variable node $i$ and a function node $a$ if and only if $i \in a$, that is the variable $s_i$ appears in $H_a$, the term of the CF associated to $a$. The CF of the whole system can then be written as:

$$H = \sum_a H_a(s_a), \text{ with } s_a = \{s_i, i \in a\} \qquad (A3)$$

Probabilistic graphical models are usually defined in a slightly different way (Smyth, 1997). In the case of *Markov random fields*, also called Markov networks, the joint distribution over all variables is given by:

$$p(s) = \frac{1}{Z}\prod_a \psi_a(s_a) \qquad (A4)$$

where $\psi_a$ is called *the potential* (potentials involving only one variable are often called *evidences*) and:

$$Z = \sum_s \prod_a \psi_a(s_a) \qquad (A5)$$

---

[7] The factor graph was used by Pelizzola (2005) in the statistical mechanics framework. There the role of cost function is played by the energy function usually called Hamiltonian.
[8] A link was considered to correspond to a correlation coefficient $|C| \geq 0.9$.

One can easily see that a combinatorial optimization model described by the cost function (A3) corresponds to a probabilistic graphical models with potentials $\psi_a = \exp(-H_a)$.

Denoting the variables as $s_1 = $ GDP, $s_2 = $ FCE, $s_3 = $ GCF and $s_4 = $ NEX, the cost function associated to the factor graph from Figure 7 is[9]:

$$H = \text{(AUT)}(s_2, s_3) + \text{(BEL)}(s_1, s_2) + \text{(DEU)}(s_1, s_2, s_4) + \text{(DNK)}(s_1, s_3) +$$
$$+ \text{(ESP)}(s_2, s_3) + \text{(FIN)}(s_3, s_4) + \text{(FRA)}(s_1, s_3) + \text{(GBR)}(s_1, s_2, s_3) +$$
$$+ \text{(IRL)}(s_1, s_2, s_4) + \text{(ITA)}(s_1, s_4) + \text{(LUX)}(s_4) + \text{(NLD)}(s_2) +$$
$$+ \text{(PRT)}(s_1, s_2, s_3, s_4) + \text{(SWE)}(s_1, s_2, s_3, s_4).$$

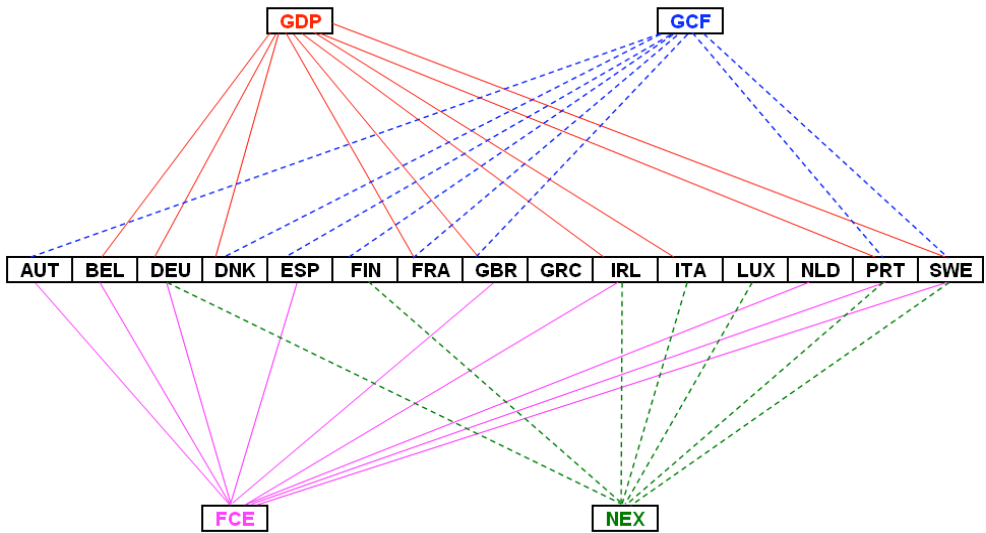

**Figure 7:** The factor graph associated to EU country connections, according to the strongest correlations extracted from Tables 2, A3, A5 and A7.

Now we define **cluster** $\alpha$ as a subset of the factor graph such that if a function node belongs to $\alpha$, then all the variable nodes $i \in a$ also belong to $\alpha$ (while the converse needs not to be true, otherwise the only legitimate clusters would be the connected components of the factor graph). Given a cluster we can define its probability distribution[10] as:

$$p_\alpha(s_\alpha) = \sum_{s \setminus s_\alpha} p(s) \qquad (A6)$$

---

[9] As Greece does not display strong correlations after the above criterion, it is not included into the cost function. If the linkage threshold is established to a lower value, *e.g.* $|C| \geq 0.8$, its function node appears as $(GR)(s_1, s_4)$, *i.e.* it belongs to the same cluster as Italy.

[10] The probability $p(s)$ is here defined as the ratio between the number of realized connections and the number of all possible connections.

and its entropy:

$$S_\alpha(s_\alpha) = -\sum_{s_\alpha} p_a(s_\alpha) \ln p_a(s_\alpha) \tag{A7}$$

Table 4 summarizes the results.

Table 4: Clustering of EU countries in a 4-variable factor graph approach

| Function Nodes | Cluster | Number of links | Number of possible links | Probability | Entropy |
|---|---|---|---|---|---|
| GDP-FCE-GCF | AUT-BEL-DNK-ESP-FRA-GBR-NLD | 14 | 28 | 0.500 | 0.347 |
| FCE-GCF-NEX | AUT-ESP-FIN-LUX-NLD | 8 | 20 | 0.400 | 0.367 |
| GDP-FCE-NEX | BEL-DEU-IRL-ITA-LUX-NLD | 12 | 24 | 0.500 | 0.347 |
| GDP-GCF-NEX | DNK-FIN-FRA-ITA-LUX | 9 | 20 | 0.450 | 0.359 |

As one can see, the maximum entropy corresponds to the clustering scheme which does *not* explicitly include GDP but its components (consumption, investments and trade), while the coupling between GDP and investments (FCE) leads to the minimal entropy clustering schemes.